\documentclass[aps,final,notitlepage,oneside,twocolumn,nobibnotes,nofootinbib,
superscriptaddress,noshowpacs,centertags]{revtex4-1}

\usepackage[english]{babel}
\usepackage{graphicx}
\usepackage{latexsym}
\usepackage{amssymb}
\usepackage{amsmath}
\usepackage{float}

\begin{document}

\title{Observational manifestations of ``cosmological  dinosaurs'' at redshifts  $z\sim20$}
\author{V. K. Dubrovich}\thanks{e-mail: dvk47@mail.ru}
\affiliation{Special Astrophysical Observatory, St. Petersburg Branch, Russian Academy of Sciences, St.
Petersburg, 196140 Russia}
\author{Yu. N. Eroshenko}\thanks{e-mail: eroshenko@inr.ac.ru}
\affiliation{ Institute for Nuclear Research of the Russian Academy of Sciences, Moscow, 117312 Russia}
\author{S. I. Grachev}\thanks{e-mail: stasgrachev47@gmail.com}
\affiliation{Sobolev Astronomical Institute, St. Petersburg State University, St. Petersburg, 198504 Russia}


\date{\today}

\begin{abstract}
We consider a primordial black hole of very high mass, $10^9-10^{10}M_\odot$, surrounded by  dark matter and barionic halo  at redshifts  $z\sim20$ without any local sources of energy release. Such heavy and concentrated objects in the early universe were previously called ``cosmological dinosaurs''. Spectral distribution and spatial variation of the brightness in the 21~cm line of atomic hydrogen are calculated with the theory of radiation transfer. It is shown that a narrow and deep absorption arises in the form of the spherical shell around the primordial black hole at the certain radius. The parameters of this shell depend almost exclusively on the mass of the black hole. The hardware and methodological aspects of the search for such objects are discussed.
\end{abstract}

\maketitle 

\tableofcontents



\section{Introduction}

The evolution of the early universe can be divided into several main stages (eras). One of such stages is the so called Dark Ages. It is characterized, in particular, by the absence of visible objects, sources of electromagnetic radiation, both in the optical and radio bands. According to the standard picture, there were small density fluctuations, expanding or, in the case of sufficiently large value, contracting against the Hubble flow due to self-gravity. Only very rare isolated objects became the first stars and protogalaxies. In addition, the entire universe was filled with cosmic microwave background radiation (CMB) having Planck spectrum and high degree of homogeneity and isotropy. This epoch follows the moment ($z\simeq1100$) of recombination, the last scattering of photon, and precedes the era ($z\sim15-20$) of plentiful star formation, called the Cosmic Dawn.
      
In the absence of intrinsic energy release, first objects can give observational manifestations only due to their interaction with CMB under the obligatory presence of some forms of nonequilibrium. An example of such a situation is the formation of spectral-spatial fluctuations of the CMB temperature due to the scattering of photons by primary molecules, where the nonequilibrium is due to peculiar velocities related to the density fluctuations  of  matter  \citep{Dubrovich77,Beretal93,DubrovichLip95}. Another type of nonequilibrium comes from the difference of the temperatures of radiation and matter arising due to the expansion of the universe \citep{VarshHers77}. In particular, it was shown by \citet{VarshHers77} that this temperature difference leads to the formation of  isotropic X-ray spectrum's  distortion  during the transfer of energy from X-ray photons to matter in the hyperfine 21~cm line of hydrogen in inelastic collisions of hydrogen atoms.
   
In this paper, unlike \citep{VarshHers77}, we consider the effects associated with the presence of a significant matter heterogeneity and the presence of peculiar velocities \citep{Dubrovich18}. Specifically, we consider the heterogeneity associated with the heavy primordial black holes (PBHs). The presence of very massive PBHs moves us beyond the standard cosmological picture of Dark Ages and allows to give some predictions that can be verified at direct astronomical observations in the foreseeable future. We point out the possibility of observing a special class of objects by their manifestation in the Dark ages era. These nonlinear gravitationally-bounded objects were called ``cosmological dinosaurs'' \citep{DubrovichGla12}. They are PBHs of sufficiently large mass, surrounded by dark matter (DM) with baryons captured. According to modern concepts, the existence of such ``dinosaurs'' is not excluded, at least in the amount that is allowed by the known observational restrictions. Given the great importance of the possible detection of such objects for fundamental physics and cosmology, a detailed description of their possible observational manifestations seems relevant.
  
The choice of the redshift at which we propose to search for these objects is determined by the following arguments:

\begin{itemize}

\item the temperature difference between matter and radiation decreases with increasing $z$;

\item at $z<15$, the reionization begins, which lowers the fraction of neutral hydrogen and, accordingly,  the optical depth in the 21~cm line decreases;

\item in addition, in the era of reionization, numerous new distortion-generating mechanisms begin to operate that are highly model-dependent and therefore significantly confuse the results.

\end{itemize}

Therefore, we chose the redshifts near $z\sim20$, where the possibility for detecting such objects is optimal.

The central object of a cosmological dinosaur is PBH. Starting with the pioneering work of \citet{ZelNov66}, as well as \citet{Haw71}, where the principle possibility of PBHs formation was described, a number of other possible scenarios were proposed. B.~Carr and S.~Hawking investigated the mechanism of PBHs formation from adiabatic density perturbations.  The formation of PBHs in the early dusty stages was also possible \citep{KhlPol80}. In \citep{BerKuzTka83,Khletal98,RubKhlSak00} another mechanism of PBH and PBHs clusters formation from domain walls has been investigated.  \citet{DolSil93} developed the scenario of PBHs formation from the perturbations of baryon charge. In recent years, PBHs have attracted increased attention because the merge of double PBHs could explain some of the gravitational-wave LIGO/Virgo events  \citep{Naketal97,DolPos20}. But all events cannot be explained by the merger of the PBH, because there are events with neutron stars. The possibility was considered that PBHs represent all or some part of the DM. 

The hypothesis that galaxies or some subclass of them can form around the PBHs as seeds was proposed by \citet{Rya72} and \citet{CarRee84}. In this paper, we also consider the formation of DM and baryonic gas objects around the PBH, which may represent some subclass of rare spheroidal galaxies today. But we propose to search for these objects at $z\sim20$, when they could produce some characteristic absorption features in the line of neutral hydrogen, available for observation by radio telescopes planned for building in the coming years.

Let us consider the PBH formed at the cosmological stage of radiation dominance. Before the dust-like phase, the fall of DM and, especially, baryons on PBHs is ineffective. The DM peaks form only in the immediate vicinity of PBHs. However, after the matter-radiation equality, the secondary accretion process \citep{Gott75,Gunn77} begins to work efficiently, and a universal DM density profile form in the  self-similar manner \citep{FilGol84,Ber85}. The growth of these halos ends when the usual density perturbations enter the nonlinear stage, which become competing centers of attraction stopping its flow of DM onto PBH \citep{DokEro01,DokEro03}.

The behaviour of baryon gas around a PBH is more complicated and depends on the mass of the PBH and the epoch under consideration. At the early stages, baryons cannot fall onto a low-mass PBH due to the gas pressure. The formation of baryon condensations begins from the Jeans mass, which depends on time and the mass of the central PBH. The trapped gas in the virial region warms up and begins to radiate. In this article, we consider the distant peripheral part of the condensation of matter around the PBH, where the density of the gas was only slightly increased compared to the average cosmological density. As it was shown by \citet{Dubrovich18,DubrovichGra19}, strong absorption in the 21~cm line is possible in this region. This absorption makes it possible in principle to detect sufficiently large nonlinear objects using radio telescopes.

We first describe in more detail the structure of cosmological dinosaurs, their composition and evolution in time, and then evaluate the possibility of their observations with the next-generation radio telescopes. We show that these objects are potentially observable with corresponding large telescopes, \citep{SKA-1}. In particular, a promising radio telescope on the far side of the moon is being considered by  \citet{SKA-2}.

\section{Structure of ``cosmological dinosaurs''}


   \subsection{Restrictions on the very massive PBH}
   
Before describing the structure of ``cosmological dinosaurs'', we briefly indicate how many seed PBHs can be in the universe.

Current observational restrictions on the number of PBHs in the universe in different mass ranges can be found in the review of \citet{Caretal20}. The restrictions are given in the form of the upper limits on the fraction $f_{\rm PBH}$ of PBHs in the composition of DM. In the range of interest $10^9-10^{10}M_\odot$ there are restrictions on the $\mu$-perturbations of CMB, millilenzing of compact radio sources, as well as a number of dynamic restrictions: on the destruction of globular clusters, heating of the stellar disk (i.e., an increase of the stars velocity dispersion), tidal destruction of galaxies, and the dynamic friction of PBHs in the halo.

The strongest constraint may come from the $\mu$-distortions of CMB. It is associated with the Silk effect and the dissipation of density perturbations. However, as indicated by \citet{Caretal20}, the $\mu$-restriction is valid only if the PBHs formed from high density peaks in the Gaussian perturbations, as in the historically first model of PBHs formation. But in other models, this restriction is not applicable. It does not work also in the case of non-Gaussianity, which is expected in the limit of high peaks. Thus, the restriction on $\mu$-perturbations of CMB is model dependent and may not be satisfied.

In this paper, we assume that the $\mu$-restriction does not play a role, and we accept as the most conservative constraint $f_{\rm PBH}\leq10^{-5}$ followed from the dynamical effect, see Fig.~13 in \citet{Caretal20}. Let us estimate the average distance between objects in this case.
Their density
\begin{equation}
n_{\rm PBH}(z)=\frac{\rho_{c,0}\Omega_m f_{\rm PBH}(1+z)^3}{M_{\rm PBH}},
\end{equation}
where $\rho_{c,0}$ is the critical density of the universe, $\Omega_m$ is the cosmological parameter of DM.
The average distance between PBHs $l(z)=(n_{\rm PBH}(z))^{-1/3}$ is
\begin{equation}
l(z)\simeq2\left(\frac{1+z}{21}\right)^{-1}
\left(\frac{f_{\rm PBH}}{10^{-5}}\right)^{-1/3}
\left(\frac{M_{\rm PBH}}{10^{9}M_\odot}\right)^{1/3}~\mbox{Mpc.}
\end{equation}
In the modern universe at $z=0$ the distance between the objects is $\geq40$~Mpc, i.e., they are quite rare.


   \subsection{PBH and the halo of dark matter}

Let us first describe the DM structure of ``cosmological dinosaurs.'' One of the characteristic scales, $r_c$,  is the distance at which the gravity of PBH predominates in the moment of equality ($z=z_{\rm eq}\simeq3119$). Consider an PBH with mass $M_{\rm PBH}$, and let $\rho_{\rm eq}$ be the density of DM at equality, then
\begin{equation}
r_c=\left(\frac{3M_{\rm PBH}}{4\pi\rho_{\rm eq}}\right)^{1/3}=50\left(\frac{M_{\rm PBH}}{10^9M_\odot}\right)^{1/3}~\mbox{pc.}
\end{equation}
The DM density peak formed inside the radius $r_c$ at the stage of radiation domination, as well as a cloud of hot ionized gas.

\begin{figure}
\centering
\includegraphics[width=\hsize]{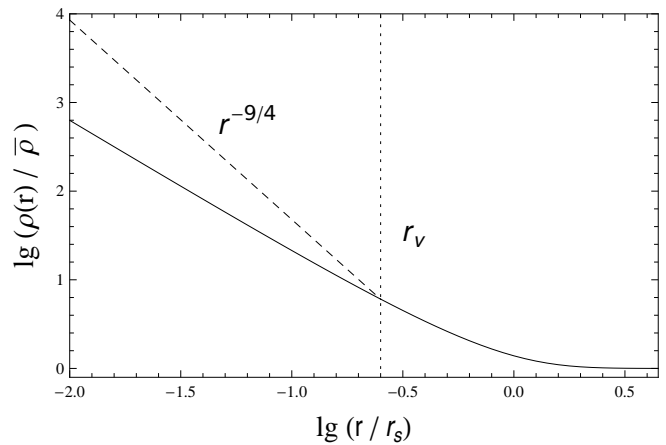}
\label{figrho}
\caption{The self-similar density profile obtained by \citet{Ber85} is shown by the solid line. The density of DM around the PBH is given in units of average cosmological density. The layer stops at the radius $r=r_s(z)$. The vertical dotted line indicates the virial radius $r_v$. To the left of this line there is a virialized mixed region with density (\ref{rhovir}), to the right there is the region of a single-stream flow. The dashed line shows the actual density in the virialized region, which differs from the density given by the self-similar solution.}
\label{auto}
\end{figure}

The next characteristic scale is the radius of the stopped DM layer  \citep{Ber85}
\begin{equation}
r_s(z)=6.7\times10^{22}\left(\frac{M_{\rm PBH}}{10^9M_\odot}\right)^{1/3}\left(\frac{1+z}{18}\right)^{-4/3}~\mbox{cm}.
\end{equation}
Inside the virial radius $r_v\simeq r_s/4$ virilization and mixing of DM occurred. In this region there is a stationary distribution of DM with the density profile obtained in the secondary accretion model
\begin{equation}
\rho(r)=2.3\times10^{-17}\left(\frac{M_{\rm PBH}}{10^9M_\odot}\right)^{3/4}\left(\frac{r}{1\mbox{~pc}}\right)^{-9/4}~\mbox{g~cm$^{-3}$}.
\label{rhovir}
\end{equation}
The density at $r>r_v$ is given by the self-similar solution found by \citet{Ber85}. This density is shown at Fig.~\ref{auto}, where the self-similar density profile is joined with the virial density profile (\ref{rhovir}).
In the self-similar solution, the DM moves to the center and is absorbed by the black hole, but in the real case, the DM is virialized and accumulated in the halo around the black hole, so the real density is higher.

   \subsection{Baryonic halo}

At the pre-recombination stage, when the radiation density is $\rho_{\rm r}$, the PBH's gravity dominates inside the radius $[3M_{\rm PBH}/(4\pi\rho_{\rm r})]^{1/3}$. However, as shown by \citet{CarRee84}, ions in this epoch not only move under the influence of the PBH's gravitational field, but they also experience friction due to scattering of photons. The balance of these forces gives the characteristic radius $r_{dr}$, inside which the gravitational force prevails, and the density of baryons can increase. Inside this radius, the characteristic particle velocities are $v(r)=(GM_{\rm PBH}/r)^{1/2}$, and the gas temperature
\begin{equation}
T(r)=\frac{G[M_{\rm PBH}+M(r)]m_p}{3rk_B},
\label{temp}
\end{equation}
where $m_p$ is the mass of the proton, $k_B$ is the Boltzmann constant. This temperature can be very high $\sim10^7-10^8$~K.

The density in this plasma clump at some arbitrary radius $r$ is of the order of the average cosmological density at that time $t(z)$ when $r_{dr}(z)=r$. Outside the radius $r_{dr}$, the density of baryons does not increase. The increase becomes possible only after recombination. As a result, at the epoch before recombination, a dense hot plasma clump with radius $r_{dr}$ is formed around the PBH. The outer boundary of the plasma clump has a radius of $\sim35$~pc in the case of the seed PBH mass $5\times10^{9}M_\odot$. The choice of the seed mass $5\times10^9M_\odot$ is due to the fact that a DM halo with a mass  $\sim10^{11}M_\odot$ is formed around such an PBH at redshift $z\sim17$. These are the objects which are most suitable for the role of ``cosmological dinosaurs.''

Let us now consider the era after recombination. At rather late times $z<150$ the baryon gas ceased to exchange heat with CMB, and its compression  is adiabatic with good accuracy. We consider scales that are several orders of magnitude greater than the Jeans radius for baryons. In this case, it is possible to neglect the pressure of the baryon gas and assume that the motion of the baryons is not delayed relative to the DM. The density of baryons in this case is $\rho_b(r)=\Omega_b\rho(r)/\Omega_m$. In addition, we consider the radii $r\gg r_c$, where the mass of dark matter exceeds the mass of the black hole, but the PBH gave the seed for the growth of the density perturbation in the homogeneous DM medium. 

The total DM mass inside a radius $r$ is given by the integral $M(r)=\int_0^r\rho(r)4\pi r^2dr$. For $r\sim r_s$, this mass is one and a half orders of magnitude greater than $M_{\rm PBH}$. In the region $r<r_v$, the virial velocity is $v=(GM/r)^{1/2}$. The temperature of baryonic gas is determined by the Eq.~(\ref{temp}). Numerically
\begin{equation}
T(r)=3\times10^6\left(\frac{M_{\rm PBH}}{10^9M_\odot}\right)^{3/4}\left(\frac{r}{1\mbox{~pc}}\right)^{-1/4}~\mbox{K}.
\end{equation}
This expression is applicable for $r_c\ll r<r_v$.

\begin{figure}
\centering
\includegraphics[width=\hsize]{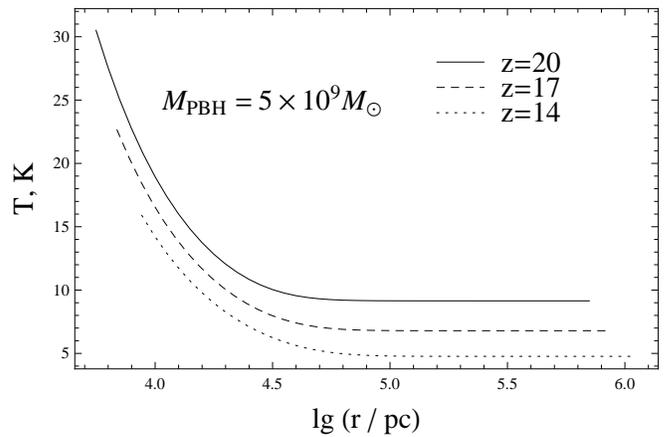}
\caption{Gas temperature around $5\times10^9M_\odot$ PBH  in the region outside the virial radius for various redshifts $z$. Solid line corresponds to $z=20$, dashed line -- to $z=17$, and dotted line -- to $z=14$.}
	\label{figtemp5x109}
\end{figure}

Consider the region $r>r_v$. To calculate the baryon gas temperature, one can use the adiabatic law $T\rho_b^{1-\gamma}=const$. Knowing the gas temperature far from the PBH, it is easy to calculate the temperature increase in the region $r>r_v$, taking into account the gas compression $\rho/\bar \rho$ shown at Fig.~\ref{auto}. For $\gamma=5/3$ one has
\begin{equation}
T=\bar T\left(\frac{\rho}{\bar \rho}\right)^{2/3},
\label{gasmom}
\end{equation}
where $\bar T$ is the temperature of gas far from PBH (for example, $\bar T=6.8$~K at $z=17$). The calculated  temperature around $5\times10^9M_\odot$ PBH  is shown at Fig.~\ref{figtemp5x109}.
In the halo regions with different densities, the gas temperature will be different. However, due to interatomic collisions in the gas, temperature equalization is possible on some scales. The calculation shows that, on the scales under consideration, the temperature equalization time exceeds the current cosmological time. Therefore, heat transfer is weak, and baryon condensations are not isothermal, but they are better described by adiabatic law (\ref{gasmom}).


\section{Brightness temperature in 21 cm line}

In this section, we examine the absorption characteristics of cosmological dinosaurs. It is assumed that there are no additional sources of energy release that could ionize a region of neutral hydrogen outside the virial radius. In particular, it is assumed that there is no disk or spherical accretion onto the PBH in the center of the object, which could create a powerful ionizing radiation. There are galaxies at $z=7.7$ inside the ionized gas bubbles of size $\sim1$~Mpc \citep{Til20}. The possibility of ionization and the formation of such bubbles in earlier eras remains unclear, and we assume further  that the powerful energy release at $z\sim20$ is unlikely.


   \subsection{Some relations from the theory of radiation transfer}

As an example, we calculate the distribution of brightness temperature over
frequencies and directions for the shells at $z = 20$. Our
consideration excludes the central (``hot'') region with the radius
5~kpc. The $r_0=44.8$~kpc is taken as the external radius, which corresponds to
doubled stop radius $r_s$. At $r=r_0$ the conditions in the shell are already close to
conditions in an unperturbed environment.

In the reference system, associated with the center of the shell, the radiation transport in the line $\lambda_{01}=21.11$~cm ($\nu_{01}=1420.4$~MHz) arises
due to transitions between hyperfine sublevels of the atomic hydrogen ground state. Let $n_0$ and $n_1$ be the populations of the lower
and upper sublevels, and $n_{\mbox{\scriptsize H}}=n_0+n_1$. We introduce the
spin temperature $T_s$: $n_1/n_0=(g_1/g_0) \exp(-T_*/T_s)$, where $g_0=1$,
$g_1=3$ are the statistical weights of the sublevels, and $T_*=h\nu_{01}/k=0.068$~K.
Next, we introduce the dimensionless frequency $x=(\nu-\nu_{01})/\Delta\nu_D(T_g)$, where
$\Delta\nu_D(T_g)=\nu_{01}u/c$, $u=\sqrt{2kT_g/M}$, and the coefficient profile
absorption $\phi(x)$ with normalization $\int_{-\infty}^{+\infty}\phi(x)dx=1$. At
the following calculations the Doppler profile $\phi(x)=(1/\sqrt{\pi})e^{-x^2}$ is used.

Large-scale motions of gas are described by the velocity field $\bf v(\bf r)$, and it is necessary to take into account
the corresponding offset of the center frequency, replacing $x$ with $x-{\bf v}\cdot{\bf n}/u$, where $\bf n$ is the orth in the direction
of radiation propagation. For a purely radial velocity field
${\bf v}=v(r){\bf r}/r$ and ${\bf v}\cdot{\bf n}/u=v(r)\mu/u$, where $\mu=
{\bf n}\cdot{\bf r}/r$. It should be noted that the dimensionless value $x$ depends on the coordinates: $x(r)\propto 1/\sqrt{T_g(r)}$. We introduce
the constant (coordinate independent) dimensionless frequency
$x=(\nu-\nu_{01})/\Delta\nu_D(T_c)$, where $T_c$ is some fixed
temperature ($T_c=236.8$~K was assumed in the calculations). Then $x(r)=x\delta(r)$,
where $\delta(r)=\sqrt{T_c/T_g(r)}$

\begin{figure}
\centering
\includegraphics[width=\hsize]{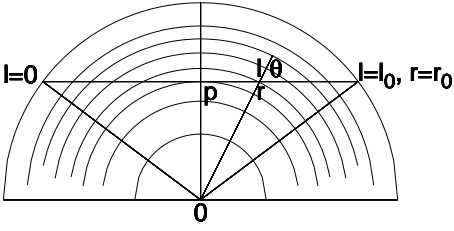}
\caption{Discretization scheme ($p$,$l$) used to solve the equation of
radiation transfer in spherical geometry.}
	\label{sg1}
\end{figure}
Consider the change of the radiation intensity $i(x,l)$ on the path element
$dl$ along the beam at an impact distance $p$ from the center of the shell. According to
Fig.~\ref{sg1} the following relations are valid
\begin{equation}
l=\mu r+\sqrt{r_0^2-p^2}, \quad p=r\sqrt{1-\mu^2},
\label{lprm}
\end{equation}
where $\mu=\cos\theta$. In the approximation of two-level atoms and under the assumption of
complete redistribution over the frequency during scattering in the line, the transport equation takes the form
\begin{equation}
\frac{di(x,l)}{dl}=-\alpha(r)\phi(x\delta(r)-\mu(r)v(r)/u(r))[i(x,l)-s(r)],
\label{difur}
\end{equation}
where $i(x,l)=(c^2/2h\nu^3)I(\nu,l)$ and $s(r)$ are dimensionless intensities
and function of sources, respectively,
\begin{equation}
\alpha(r)=\alpha_0(r)(1-e^{-T_*/T_s(r)}),
\label{asa0}
\end{equation}
\begin{equation}
s(r)=(e^{T_*/T_s(r)}-1)^{-1},
\end{equation}
\begin{equation}
\alpha_0(r)=\frac{3}{8\pi u(r)}\lambda_{01}^3A_{10}n_0(r),
\end{equation}
and $A_{10}=2.85\cdot 10^{-15}$~s$^{-1}$ is the Einstein coefficient of
spontaneous transition probability.


   \subsection{Formal solution and its properties}

Formal, i.e. for a given function of sources $s$, the solution of the equation
(\ref{difur}) has the form
\begin{eqnarray}
&&i(x,l)=i_0(x) e^{-\tau_l(x)}+
\label{fsol}
\\
&&\int_0^l e^{-[\tau_l(x)-\tau_{l'}(x)]}s(r')
\phi(x\delta(r')-\mu(r')v(r')/u(r'))\alpha(r')dl',
\nonumber
\end{eqnarray}
where $i_0(x)\equiv i(x,0)$ is given by the boundary condition (for $l=0$) and
\begin{equation}
\tau_l(x)=\int_0^l\phi(x\delta(r')-\mu(r')v(r')/u(r'))\alpha(r')dl'
\label{taul}
\end{equation}
is the optical distance along the beam at a frequency $x$.

The formal solution (\ref{fsol}) should be supplemented by the equation of
statistical equilibrium  
\begin{equation}
n_1A_{10}(1+j)+n_1 n_{\mbox{\scriptsize H}} q_{10}=\frac{g_1}{g_0}A_{10}n_0
j+n_0 n_{\mbox{\scriptsize H}}q_{01},
\label{steq}
\end{equation}
where the intensity averaged over  frequencies and directions is
\begin{equation}
j(r)=\frac{1}{2}\int_{-\infty}^{+\infty}dx\int_{-1}^{1}\phi(x\delta(r)-v(r)
\mu/u(r))i(x,l)d\mu,
\label{jeq}
\end{equation}
and $q_{01}$, $q_{10}$ are the probability coefficients of shock transitions,
related by the equation $q_{01}=q_{10}(g_1/g_0)\exp(-T_*/T_g)$. Here
the shock transitions with hydrogen atoms are taken into account, since they
play a major role in the problem under consideration. We use the table of
coefficients $q_{10}$ from the review article of \citet{FurOhBri06}. Given the relationship between the coefficients, the (\ref{steq}) is rewritten in
form
\begin{equation}
s(r)=\lambda(r) j(r) + [1-\lambda(r)]b(T_g),
\label{sleq}
\end{equation}
where $b(T_g)=(e^{T_*/T_g}-1)^{-1}$, and
\begin{equation}
\lambda(r) = \frac{A_{10}}{A_{10}+n_{\mbox{\scriptsize H}} q_{10}
[1-\exp(-T_*/T_g)]}
\label{lambda}
\end{equation}
is the probability (or albedo) of single scattering in the line.

The shell is in the field of blackbody isotropic background radiation
(CMB) at a given redshift $z$, so at the cloud boundary $r=r_0$ one has
$i_0(x)=b(T_r)\equiv 1/(e^{T_*/T_r}-1)$ for $\mu<0$. Here $T_r=T_0(1+z)$ is the
CMB temperature. For $z$ in the range 25 -- 15, the relations $T_*/T_r$,
$T_*/T_g$ and $T_*/T_s(r)$ $\ll 1$, therefore $b(T_r)\sim T_r/T_*\gg 1$,
$b(T_g)\sim T_g/T_*\gg 1$ and $s(r)=b(T_s(r))\sim T_s(r)/T_*\gg 1$. From the last inequality it follows that $n_1/n_0=g_1/g_0=3$ and, therefore,
$n_0=(1/4)n_{\mbox{\scriptsize H}}$ and $n_1=(3/4)n_{\mbox{\scriptsize H}}$.
In addition, the Eq.~(\ref{taul}) for the optical distance $\tau_l$, for which according to (\ref{asa0}) $\alpha=
\alpha_0/[1+s(r')]$, can be rewritten as
\begin{eqnarray}
\tau_l(x)=\int_0^l\phi(x\delta(r')-\mu(r')v(r')/u(r'))\frac{\alpha_0(r')}
{s(r')}dl'
\nonumber
\\
=
\int_0^l\phi(x\delta(r')-\mu(r')v(r')/u(r'))\alpha_0(r')\frac{T_*}{T_s(r')}dl',
\label{tauls}
\end{eqnarray}
and the boundary condition takes the form $i(x,0)=i_0(x)=T_r/T_*$ for $\mu<0$.
In this case, the formal solution (\ref{fsol}) and the Eq.~(\ref{sleq}) can be rewritten
respectively in the form
\begin{equation}
T_b(x,l)=T_r e^{-\tau_l(x)}+\int_0^{\tau_l(x)}e^{-[\tau_l(x)-\tau_{l'}(x)]}
T_s(r')d\tau_{l'}(x)
\label{Tbxl}
\end{equation}
and
\begin{equation}
T_s(r)=\lambda(r)\overline{T}_b(r)+[1-\lambda(r)]T_g(r),
\label{Tsr}
\end{equation}
where the brightness temperature averaged over directions and frequencies is
\begin{equation}
\overline{T}_b(r)=\frac{1}{2}\int_{-\infty}^{+\infty}dx\int_{-1}^1
\phi(x\delta(r)-\mu v(r)/u(r))T_b(x,l)d\mu,
\label{Tbmn}
\end{equation}
and $l=\mu r+\sqrt{r_0^2-r^2+r^2\mu^2}$ according to the relations (\ref{lprm}).


   \subsection{Results}

\begin{figure}
\centering
\includegraphics[width=\hsize]{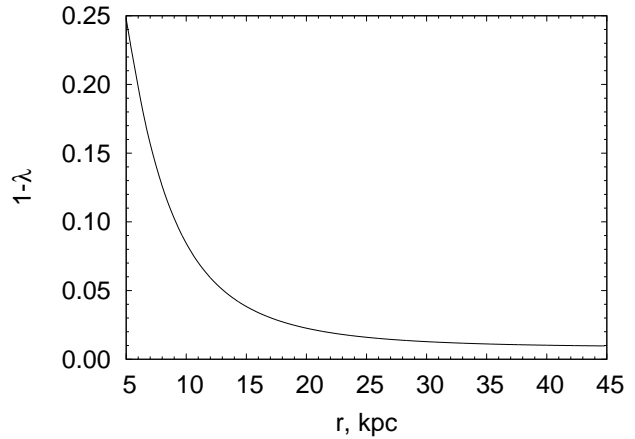}
\caption{Probability of a photon dying during a single scattering.}
	\label{sg2}
\end{figure}

\begin{figure}
\centering
\includegraphics[width=\hsize]{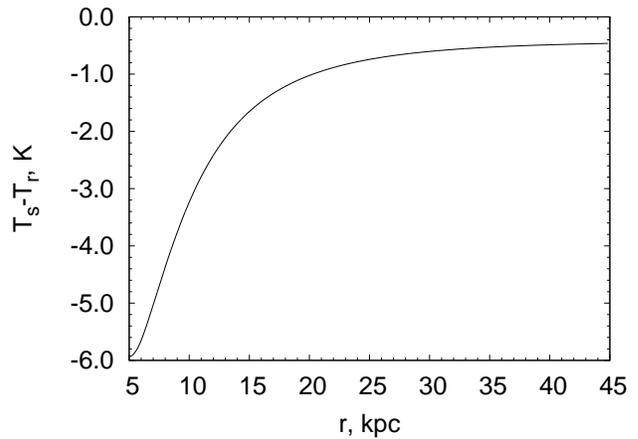}
\caption{Initial distribution of spin temperature.}
	\label{sg3}
\end{figure}

The system of equations (\ref{Tbxl}) and (\ref{Tbmn}) for determining the brightness and
spin temperature is nonlinear since spin temperature enters
into these equations nonlinearly through the optical distance (see (\ref{tauls})). However, this nonlinearity leads to the strong enlightenment of the medium
due to forced  radiative down-transitions. As a result the optical  thickness of the
shell in the center of the line is small, and the mentioned system of
equations can be solved through iterations as follows: by setting the initial
distribution of spin temperature, we calculate optical distances according to (\ref{tauls}) and then we calculate the distribution of brightness temperature 
from  (\ref{Tbxl}) and find its average value by  (\ref{Tbmn}). Further from (\ref{Tsr}) we determine the new distribution of the spin
temperature etc. These are the so-called ``lambda iterations'', or iterations by number of
scattering. Since with a small optical thickness of the shell average
brightness temperature $\overline{T}_b(r)$ is not much different from the temperature
of background radiation $T_r$, it is possible to
take the temperatures
\begin{equation}
T_s(r)=\lambda(r)T_r+[1-\lambda(r)]T_g(r),
\label{Tsr0}
\end{equation}
as the initial spin distribution, according to (\ref{Tsr}). Fig.~\ref{sg2} and Fig.~\ref{sg3} show  $1-\lambda$ and
$T_s-T_r=-(1-\lambda)(T_r-T_g)$ respectively.

\begin{figure}
\centering
\includegraphics[width=\hsize]{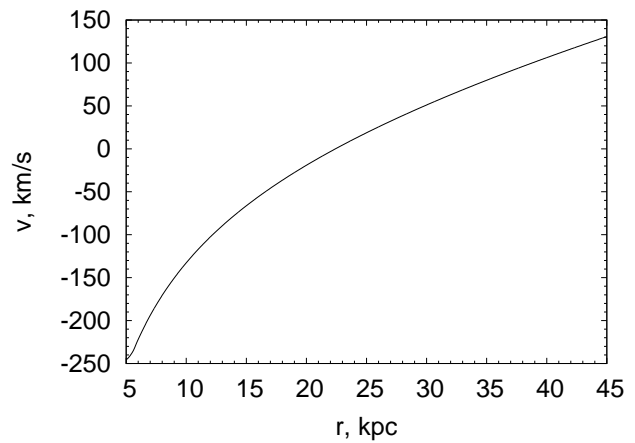}
\caption{Radial velocity distribution.}
	\label{sg4}
\end{figure}
In what follows, we restrict ourselves only by the initial approximation, which, however,
allows to identify the main features of the intensity distribution of the outgoing
radiation in directions and frequencies. The main difficulty is in the calculation of the
frequency profiles. The fact is that the range of change of
radial velocity in the shell significantly exceeds the range of variation of
thermal velocity, and the radial velocity $v(r)$ also changes sign as
crossing the point $r=r_s=22.4$~kpc (see Fig.~\ref{sg4}). This leads to
the maximum contribution into the absorption of radiation at a given fixed
frequency, gives one or more (at most three) areas on the line of sight with strongly
different physical conditions. The width of these areas is small due to the
strong dependence of the Doppler  absorption profile on the frequency.

According to (\ref{tauls}) these are areas
around points (let's call them resonant) at which the argument of the 
absorption coefficient vanishes: $x\delta(r)-\mu v(r)/u(r)=0$. From this condition we get
$\delta(r)[x-\mu v(r)/u_c]=0$, where $u_c=\sqrt{2kT_c/Mc^2}$ as defined
by the dimensionless frequency scale $x$ given above. As a result, we have the equation
$x-\mu v(r)/u_c=0$ to determine the positions of the resonance points. However instead
of search for resonance points for a given set of frequencies, it is easier to find
resonant frequencies for a given ray sampling grid ($l_k$) and
impact distances ($p_i$): $x_{ki}=\mu_{ki} v_{ki}/u_c$.
Discretization is performed as follows (see Fig.~\ref{sg1}): a set of
radial distances $r_i$ (they are the impact distances $p_i$ for rays), where
$i=1,2 ... I$, and the points $l_k$ along the ray are the points of intersection of this ray with
circles. Their number on the $i$th ray is $K=2(I-i)+1$. Calculations
were carried out for $I=2000$.

\begin{figure}
\centering
\includegraphics[width=\hsize]{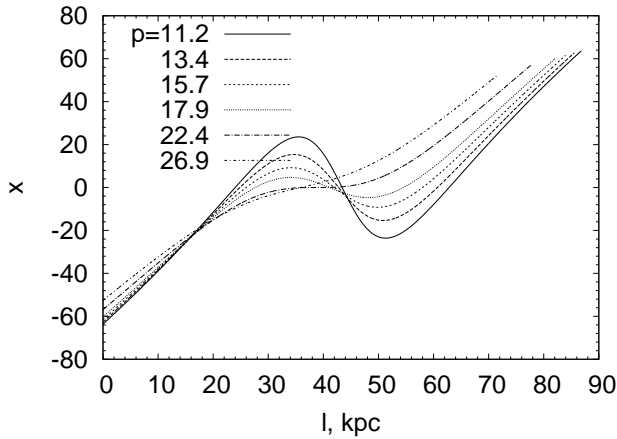}
\caption{Resonance frequency distributions for beams with different
impact distances $p$ (in kpc).}
	\label{sg5}
\end{figure}
\begin{figure}
\centering
\includegraphics[width=\hsize]{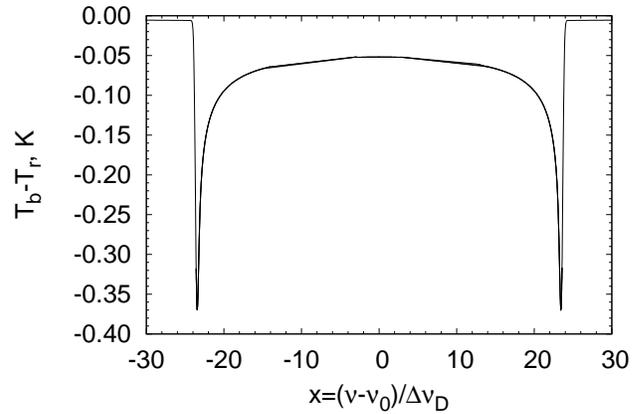}
\caption{Frequency profile of brightness temperature at impact distance
$p=r_s/2=11.2$~kpc.}
	\label{sg6}
\end{figure}
Fig.~\ref{sg5} shows the distribution of resonant frequencies for rays with
different impact distances. Here the beam with impact distance
$p=r_s=22.4$~kpc is tangent to the surface of zero radial velocity.
As can be seen, the rays with $p<r_s$ crossing this surface can contain
three regions contributing to the absorption of radiation at the same frequency $x$.
In addition, these rays have symmetrically located points of local
maximum and minimum resonant frequencies $x$, in which $dx/dl=0$. But according to
the definition of the resonant frequency, this means that at these points the derivative
of radial velocity in the direction of the beam is zero:
\begin{equation}
\frac{dv_l}{dl}=\mu^2\frac{dv(r)}{dr}+(1-\mu^2)\frac{v(r)}{r}=0.
\end{equation}

\begin{figure}
\centering
\includegraphics[width=\hsize]{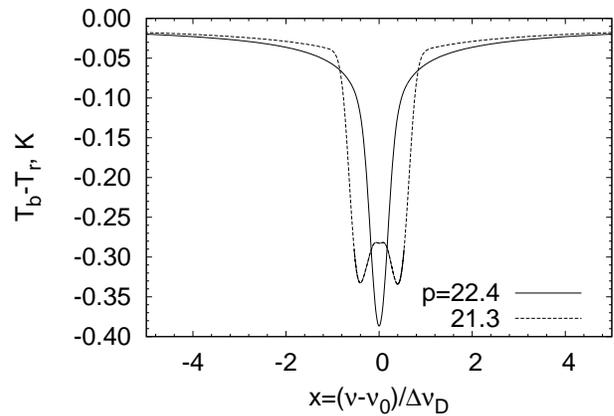}
\caption{The same as in the previous figure, but for impact distances
$p= 21.3$ and $p=r_s=22.4$~kpc.}
	\label{sg7}
\end{figure}

Here, the enlightenment of the medium does not occur due to the absence of velocity gradient,
and therefore at symmetrically positioned (relative to zero)
corresponding resonant frequencies narrow deep
absorption components should be observed, which one can see at Fig.~\ref{sg6} near $x=\pm23.6$~kpc.
With an increase of the impact distance of the beam, these components shift to the center of
the lines and eventually merge into one narrow absorption component on the beam with
impact distance equal to the stop radius: $p=r_s=22.4$~kpc (see Fig.~\ref{sg7}).

\begin{figure}
\centering
\includegraphics[width=\hsize]{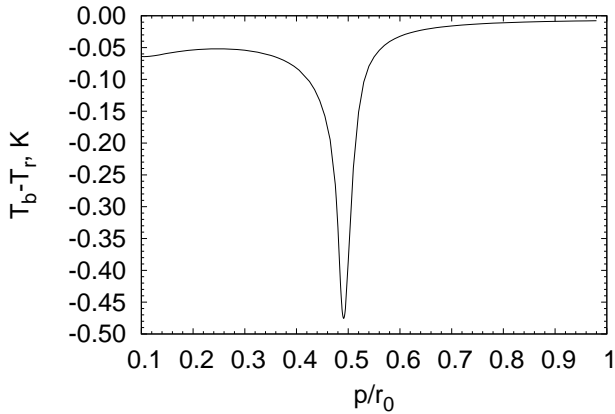}
\caption{Distribution of brightness of the disk at a frequency of $x=0$ ($r_0=2r_s=44.8$~kpc).}
	\label{sg8}
\end{figure}
Fig.~\ref{sg8} shows the distribution of brightness temperature over the disk at
the center frequency $x=0$. General slight decrease in brightness when switching from
the edge of the disk toward the center is determined by an increase in the density of the substance, as well as
decrease of spin temperature (see Fig.~\ref{sg3}) caused by an increase in probability
of photons' death in the line $1-\lambda$ (see Fig.~\ref{sg2}), which, in turn,
determined by a sharp increase in the deactivation coefficient $q_{10}$ of the upper
transition level with increasing gas temperature in the transition from the edge of the shell to
to the center. The last effect determines, in particular, the line depth in the case of
static shell, as was shown by us in the previous work of \citet{DubrovichGra19}. A sharp decrease in brightness in a narrow area near
$p/r_0=0.5$ ($p\approx 22$~kpc) is caused by strong absorption of radiation on
line of sight at distances from the center close to the radius of the stop
$r_s=22.4$~kpc, where the radial velocity gradient is zero.


\section{Conclusion}

The first light sources, stars and galaxies, form effectively starting from redshifts $z\leq10-15$. The earlier era is called the Dark Ages, because it is usually believed that luminous objects absent at that time. It is extremely difficult to study physical processes in the Dark Ages because of the absence of luminous matter. At the same time, this era is of great interest, because the first gravitationally bound objects formed and the reionization of the universe began.  The direct dynamic tests are applicable for the study of relatively close cosmological objects (for example, \citep[see][]{RaiPopOrl18}), but the development of new approaches is required to study the Dark Ages. This article discusses a new class of observations -- the observation of absorption on the individual objects in 21~cm line at redshifts of $z\sim20$. We show that despite the sophistication of this task, it can be feasible in the coming years with the developing $1$~km$^2$ radio telescopes.

At redshifts $z>15-20$, the majority of objects were still at the linear stage of their evolution, and only much later they evolved into stars and galaxies. However, in the rare DM halos (at the tail of the Gaussian distribution of perturbations), nonlinearity was reached in this era and the first stars began to appear. This is the standard picture. But it is possible that the first nonlinear objects appeared much earlier, even at the cosmological stage of radiation domination. The example of such objects are PBHs surrounded by the DM halo. The masses of these objects can be, in principle, any, and their undetectability in the modern universe can be due to the rarity of such objects and their low visibility due to the lack of strong accretion onto a black hole.  These  very rare objects are called ``cosmological dinosaurs''. In this article, we discuss the possibility of observing them with the help of radio telescopes of next generation, by their manifestation in the era of Dark Ages. Cosmological dinosaurs are PBHs of a sufficiently large masses ($M_{\rm PBH}\geq10^5-10^6M_\odot$), around which a halo of dark matter was formed and the baryons were captured. It was previously shown by \citet{DubrovichGra19} that such objects are surrounded by a region of strong absorption in 21~cm line. The radio telescopes with an area of more than $1$~km$^2$ would allow to search for such objects.

In this article we discussed the following main points:

\begin{itemize}

\item Around a PBH of large mass at the pre-stellar stage a specific picture of the background brightness field is formed in the 21~cm line  (Fig.~\ref{sg8});
 
\item  the absorption leads to the sharp increase in the optical depth in 21~cm line at the expansion stop \citep{Dubrovich18,DubrovichGra19};
 
\item the position relative to the center and the parameters of the absorption region depend only on the mass of the central PBH;

\item this picture is practically independent of the details of the processes of matter movement and energy release during accretion inside the virialization sphere;
 
\item thus, the most observed effect is mandatory for PBH;
 
\item the angular size of the absorption ring is quite small and therefore, telescopes of the SKA type \citep{SKA-1}  are necessary for its observation.

\end{itemize}

\begin{acknowledgements}
The work was performed as part of the government contract of the SAO RAS approved by the Ministry of Science and Higher Education of the Russian Federation.  
\end{acknowledgements}

\bibliography{refs}

\end{document}